\documentclass[pre,showkeys,preprintnumbers,amsmath,amssymb,superscriptaddress,twocolumn]{revtex4}
\usepackage{graphicx}
\usepackage{color}
\usepackage{bm}
\usepackage{ifthen}

\newcounter{highlighting}
\def\B{{\mathcal B}}
\def\R{{\mathbb R}}

\def\ctanh{\mathrm{ctanh}}
\def\ve{{\varepsilon}}
\def\x{\bm{x}}

\begin{document}

\title{Steady-state reaction rate of diffusion-controlled reactions in sheets}

\author{Denis S. Grebenkov}
  \email{denis.grebenkov@polytechnique.edu}
\affiliation{Laboratoire de Physique de la Mati\`ere Condens\'ee (UMR 7643), CNRS -- Ecole Polytechnique, University Paris-Saclay, 91128 Palaiseau, France}

\author{Diego Krapf}
  \email{diego.krapf@colostate.edu}
\affiliation{Electrical and Computer Engineering, Colorado State University, Fort Collins, CO 80523, USA}
\affiliation{School of Biomedical Engineering, Colorado State University, Fort Collins, CO 80523, USA}

\date{\today}

\begin{abstract}
In many biological situations, a species arriving from a remote source
diffuses in a domain confined between two parallel surfaces until it
finds a binding partner.  Since such a geometric shape falls in
between two- and three-dimensional settings, the behavior of the
macroscopic reaction rate and its dependence on geometric parameters
are not yet understood.  Modeling the geometric setup by a capped
cylinder with a concentric disk-like reactive region on one of the
lateral surfaces, we provide an exact semi-analytical solution of the
steady-state diffusion equation and compute the diffusive flux onto
the reactive region.  We explore the dependence of the macroscopic
reaction rate on the geometric parameters and derive asymptotic
results in several limits.  Using the self-consistent approximation,
we also obtain a simple fully explicit formula for the reaction rate
that exhibits a transition from two-dimensional to three-dimensional
behavior as the separation distance between lateral surfaces
increases.  Biological implications of these results are discussed.
\end{abstract}



\maketitle

\section{Introduction}

Diffusion is omnipresent in biological systems.  In particular, the
random motion of ions and molecules in aqueous environments is a
critical mechanism responsible for bringing reactants to their
reaction centers.  Without the effect of actively-driven processes,
the concentration $C$ of a species moving with diffusivity $D$
satisfies the diffusion equation 
\begin{equation*} 
\frac{\partial C}{\partial t} = D \Delta C, 
\end{equation*}
where $\Delta = \nabla^2$ is the Laplace operator.  One should note
that the diffusion equation is limited to the case of homogeneous
purely viscous liquids and thus it often breaks down in live cells as
well as in other complex fluids
\cite{bouchaud1990anomalous,klafter2005anomalous,guigas2008sampling,jeon2011vivo,sokolov2012models,metzler2014anomalous,krapf2015mechanisms},
but it is typically an excellent approximation for the motion of small
molecules in three dimensions within the range of biologically
relevant experimental times \cite{Elowitz99,Verkman02}.  At
intermediate time scales, the concentration of most species is often
found in steady state and, thus, the diffusion equation simplifies to
the Laplace equation 
\begin{equation*}
\Delta C = 0.  
\end{equation*}
In spite of the relative simplicity of this equation, solving it can
be non-trivial due to the complex morphology of cellular environments
\cite{Carslaw,Redner,Benichou10,Benichou14}.

Among the different morphologies that appear within cells, the
occurrence of thin interconnected sheets is widespread.  For example,
two cellular organelles involve the presence of aqueous environments
within thin sheets: mitochondria and the endoplasmic reticulum
(ER). On one hand, a mitochondrion contains an outer and an inner
membrane that separate this organelle into distinct compartments with
different functions, namely an intermembrane space, cristae formed by
foldings within the inner membrane and a matrix enclosed by the inner
membrane \cite{frey2000internal}.  The intermembrane space is a sheet
between the outer and inner membranes of approximately 8 nm in
thickness \cite{nicastro2000cryo,van2009tools}.  This compartment has
multiple physiological (including its role in oxidative
phosphorylation) and pathological functions, and many proteins
involved in mitochondrial signaling pathways are specifically targeted
to it \cite{van2003mitochondrial,griparic2004loss}.  On the other
hand, the ER is a continuous membrane system with a common enclosed
space comprising an intricate three-dimensional network
\cite{friedman2011er,joensuu2014er,shibata2010mechanisms}.  The ER
lumen, i.e., its interior, is filled with ions, small molecules, and
proteins.  In animal cells, the ER is the primary storage site for
intracellular Ca$^{2+}$ that can be released as Ca$^{2+}$ signals
\cite{soboloff2011sensing}.  Diffusion within the ER lumen is
essential for critical cellular processes including protein transport
and posttranslational regulation, and quantitative diffusion
measurements of Ca$^{2+}$ and proteins therein have been reported
\cite{dayel1999diffusion,verkman2002solute,okubo2015visualization}.
The peripheral ER consists of sheets and a network of tubules, where,
in mammals, ER sheets are typically in the range of 50 nm in thickness
\cite{terasaki2013stacked}.

Besides their occurrence within cell organelles, sheet-shaped
structures are common in extracellular spaces where signaling between
adjacent cells takes place.  These structures are notably prevalent in
different brain regions where communication between cells and
regulation of extracellular components is of utmost importance for
cognitive functions.  In particular, the concentrations of K$^{+}$ and
neurotransmitters such as glutamate are tightly regulated in the
extracellular space surrounding neurons. Glutamate, the major
excitatory neurotransmitter in the brain of vertebrates, when present
in excess for extended periods of time, acts as a neurotoxin that
triggers cell death.  In order to remove excess glutamate, astrocytes
take up extracellular glutamate via glutamate transporters so that
neurotransmitters are maintained at a low concentration close to
resting cells.  The best-known region for the action of glutamate
transporters in neurotransmitter removal is the synaptic cleft, where
astrocytic membranes are observed to wrap the synapse region and
express high levels of glutamate transporters
\cite{tzingounis2007glutamate,kessler2013control}.  Besides their role
in synaptic transmission, astrocytes are also observed to regulate the
concentration of neurotransmitters in extrasynaptic and
somato-dendritic regions
\cite{herman2007extracellular,mulholland2008glutamate}. Furthermore,
astrocytic signals are triggered upon binding of glutamate to
receptors, which can result in Ca$^{2+}$ signaling and the release of
glio-transmitters like glutamate, ATP and D-serine
\cite{rose2017astroglial}. Most of the processes that regulate
glutamate are modulated by glutamate diffusion within extracellular
sheets and tunnels. These extracellular spaces are found to have a
thickness of the order of 20 nm
\cite{castejon2009,kinney2013extracellular}.
       
In this article, we model how the localization of binding partners
alters local concentration within confined sheet-like spaces such as
those encountered in the extracellular space or in the peripheral ER.
In particular, we investigate the flux associated with the clearing of
glutamate from the vicinity of neuronal glutamate receptors.  These
problems are addressed by solving the three-dimensional
reaction-diffusion equations in steady state. The diffusion-limited
solution provides an upper bound for the glutamate flux.  This is a
key aspect in understanding glutamate uptake from the extracellular
gap.

The macroscopic reaction rate $J$ of steady-state diffusion-limited
reactions has been studied over the last century
\cite{Smoluchowski17,Collins49,Samson77,Samson78,Sano79,Torney83,Berg85,Tsao01,Traytak07,Strieder09,Eun13,Biello15}. 
In the three-dimensional setting, $J$ is often estimated as the
Smoluchowski reaction rate on a spherical reactive region of radius
$\rho$,
\begin{equation}   \label{eq:Smoluchowski}
J_S = 4\pi C_0 D \rho, 
\end{equation}
where $C_0$ is the concentration of molecules at an (infinitely)
distant source \cite{Smoluchowski17}.  In turn, a steady-state
solution in two dimensions is only defined for a source at a finite
distance from the reactive region, and the rate $J$ depends on this
distance (see below, as well as the related discussion in
\cite{Torney83,Berg85}).  Since three-dimensional diffusion between
parallel sheets appears to be in between these two conventional cases,
the behavior of the reaction rate $J$ is not well understood.  The aim
of the paper is to determine the dependence of the reaction rate $J$
on the geometric parameters of the problem as such the size of the
reactive region, the distance to the source, and the separation
between sheets.

The paper is organized as follows.  In Sec. \ref{sec:model}, we
present the mathematical model and its solution, and explore the
dependence of the reaction rate on the geometric parameters of the
problem.  In Sec. \ref{sec:discussion}, we discuss some limitations of
the considered model and the related extensions, as well as the
biological implications.  Technical calculations are reported in
Appendices.

\section{Mathematical model and solution}
\label{sec:model}

\begin{figure}
\begin{center}
\includegraphics[width=40mm]{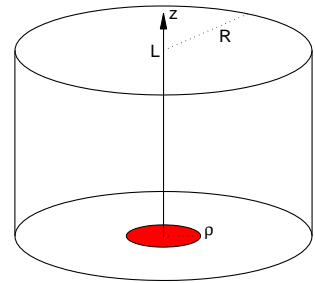} 
\end{center}
\caption{
A disk-like sink of radius $\rho$ (in red) inside the capped cylinder
of radius $R$ and of height $L$.  The source of particles is located
at the cylindrical surface.  In a typical biological setting, one has
$R \gg L \gg \rho$ so that the cylinder should be extended in the
radial (lateral) direction. }
\label{fig:scheme}
\end{figure}

We model a thin sheet (e.g., a flat junction between two cells) as a
capped cylinder of radius $R$ (the radius of junction) and of height
$L$ (the distance between cells), see Fig. \ref{fig:scheme}.  In
cylindrical coordinates $(r,z,\varphi)$, this confining domain is
described as
\begin{equation*}
\Omega = \{ 0 < r < R, ~ 0 < z < L, ~ 0 \leq \varphi < 2\pi\}.  
\end{equation*}
On one surface (at $z = 0$), there is a target protein that is modeled
by a reactive disk $\Gamma$ of radius $\rho$.  Once a molecule (e.g.,
glutamate) arrives onto the target protein, it is adsorbed and removed
from the domain.  We aim at computing the macroscopic reaction rate
$J$, i.e., the steady-state diffusive flux of particles onto the
reactive protein coming from outside of the junction.  In the steady
state, one can assume that multiple sources of particles, distributed
in the space outside the junction, maintain the concentration of
particles constant at the outer boundary of the junction (i.e., at $r
= R$).  Fixing the boundary condition at the outer boundary allows one
to disentangle the diffusion-reaction problem inside the junction from
diffusion in the outer space toward the junction.  The latter
determines only the constant concentration $C_0$ at the outer boundary
which is just a proportionality factor due to the linear character of
the problem.  In turn, the solution of the diffusion-reaction problem
inside the junction depends on the geometric parameters of the
junction: the radius $R$ and the height $L$ of the junction, as well
as the radius $\rho$ and the location of the target protein.  In this
paper, we focus on the role of these geometric factors.  We emphasize
that this geometric model is different from a model of concentric
cylinders with a reactive region on the inner cylinder that was
studied in \cite{Grebenkov17,Grebenkov18} in the context of first
passage phenomena.  In particular, in our setting, the particles reach
the reactive region from above.

In mathematical terms, one needs first to determine the steady-state
concentration of particles in the junction, $C(r,z,\varphi)$, by
solving the boundary value problem:
\begin{subequations}  \label{eq:Eq}
\begin{eqnarray}
\label{eq:Eq1}
\Delta C(r,z,\varphi) &=& 0 \quad \textrm{in the junction} ,\\
\label{eq:Eq2}
C &=& C_0 \quad \textrm{on the outer boundary} , \\
\label{eq:Eq3}
C &=& 0 \quad \textrm{on reactive region $\Gamma$}, \\
\label{eq:Eq4}
\partial_n C &=& 0 \quad \textrm{on the cell membranes},
\end{eqnarray}
\end{subequations}
where $\partial_n$ is the normal derivative directed outward the
domain.  The Dirichlet boundary condition (\ref{eq:Eq3}) at the
reactive patch expresses a reaction on the target protein upon the
first encounter.  This perfect reaction can be replaced by a partial
reaction modeled by a Robin boundary condition (see
Sec. \ref{sec:discussion}).  Finally, the Neumann boundary condition
(\ref{eq:Eq4}) ensures that the two lateral surfaces at $z = 0$ and $z
= L$, representing the cell membranes, are impenetrable to the
particles (except for the reactive region).  Once this problem is
solved, the diffusive flux onto the reactive region $\Gamma$ is
obtained by integrating the flux density over $\Gamma$:
\begin{equation}
J = \int\limits_\Gamma d{\bf s} \, ( - D \partial_n C)_{|\Gamma} \,.
\end{equation}
As we are interested in the effect of geometric parameters, it is
convenient to compare the flux to the classic Smoluchowski flux $J_S$
from Eq. (\ref{eq:Smoluchowski}):
\begin{equation}
\Psi = \frac{J}{4\pi C_0 D \rho} \,.
\end{equation}
The normalized flux $\Psi$ does not depend on the imposed
concentration $C_0$.  Moreover, there is also no dependence on the
diffusion coefficient $D$ for the considered case of perfect
steady-state reactions.

\begin{figure}
\begin{center}
\includegraphics[width=85mm]{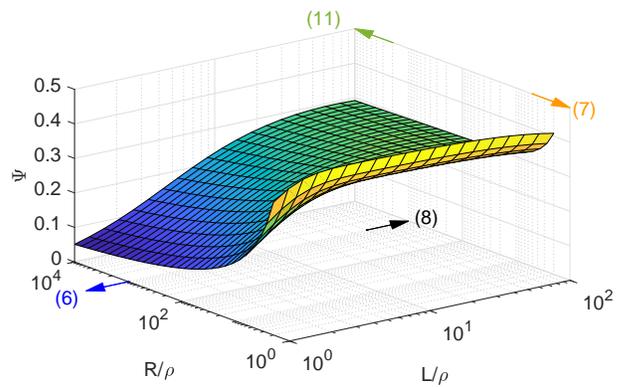}  
\end{center}
\caption{
The normalized flux $\Psi$ as a function of two geometric parameters:
$L/\rho$ and $R/\rho$.  For convenience, the practically irrelevant
region with $R/\rho \approx 1$ is excluded by setting $R/\rho \geq 2$.
The colored arrows indicate the Eq. numbers that show the derived
asymptotic relations with their corresponding limits, such as $L/\rho
\to \infty$. }
\label{fig:psi}
\end{figure}

When the target protein is located at the center of the surface of one
cellular membrane (i.e., it is concentric with the junction, see
Fig. \ref{fig:scheme}), the boundary value problem (\ref{eq:Eq}) can
be solved semi-analytically.  In fact, although the solution technique
is standard (see Appendix \ref{sec:solution} for details), the
resulting expressions for the concentration $C(r,z,\varphi)$ and for
the macroscopic reaction rate $J$ are not fully explicit, involving
the inversion of an infinite-dimensional matrix.  While this step has
to be done numerically, the computation is fast and accurate, allowing
one to explore the dependence of $\Psi$ on the two geometric ratios
$L/\rho$ and $R/\rho$.  Figure \ref{fig:psi} shows the normalized flux
$\Psi$ as a function of these parameters (note that $R/\rho \geq 1$,
whereas $L/\rho$ can range from $0$ to $\infty$).  Let us explore the
dependence on both geometric parameters.

In the regime $L \ll \rho$, the separation between two lateral
boundaries is so short that a particle appearing above the reactive
region rapidly reaches this region and reacts.  This regime is
therefore close to diffusion between two coaxial cylinders of radii
$R$ and $\rho$, for which the concentration profile and the flux are
well known:
\begin{equation}  \label{eq:J0flux}
C_{\rm cyl}(r) = C_0 \frac{\ln(r/\rho)}{\ln(R/\rho)} \,,  \qquad
J_{\rm cyl} = \frac{2\pi D L C_0}{\ln(R/\rho)} \,.
\end{equation}
As for the two-dimensional diffusion problem, these solutions vanish
logarithmically as $R\to\infty$.  Dividing this flux by $J_S$, one
gets the asymptotic behavior of $\Psi$ as $L\to 0$:
\begin{equation}  \label{eq:Psi_cyl}
\Psi \simeq \Psi_{\rm cyl} = \frac{L/\rho}{2\ln(R/\rho)} \qquad (L \ll \rho)\,.
\end{equation}

In the limit $R \to \rho$, the concentration $C(r,z,\varphi)$ can be
found in a fully explicit form but the diffusive flux diverges
logarithmically in this limit (see Appendix \ref{sec:Rrho} for
details):
\begin{equation}
\Psi \to \frac{1}{\pi} \ln \biggl(\frac{\rho}{R-\rho}\biggr) + O(1)  \qquad (R-\rho \ll \rho).
\end{equation}
This is a consequence of zero distance between the source (here, the
cylinder at $r = \rho$) and the sink (the disk at $z = 0$) that touch
each other.

In the limit $L \to \infty$, the problem is close to that of an
absorbing disk in the half-space, for which (see Appendix
\ref{sec:half-space}):
\begin{equation}  \label{eq:Psi_disk}
\Psi \simeq \Psi_{\rm disk} = \frac{1}{\pi} \approx 0.3183\ldots  \qquad (L\gg R).
\end{equation}
In this regime, the geometric confinement is irrelevant, and the
diffusive flux onto the target protein is close to the Smoluchowski
limit.  The reduction by the factor $\pi$ is due to our choice of
modeling the target protein by a disk instead of a sphere.  To fully
illustrate the specific role of the aspect ratio of the reactive
region, we compute the normalized flux $\Psi$ for an oblate spheroid,
allowing for a continuous variation from a sphere to a disk (see
Appendix \ref{sec:oblate}).

\begin{figure}
\begin{center}
\includegraphics[width=85mm]{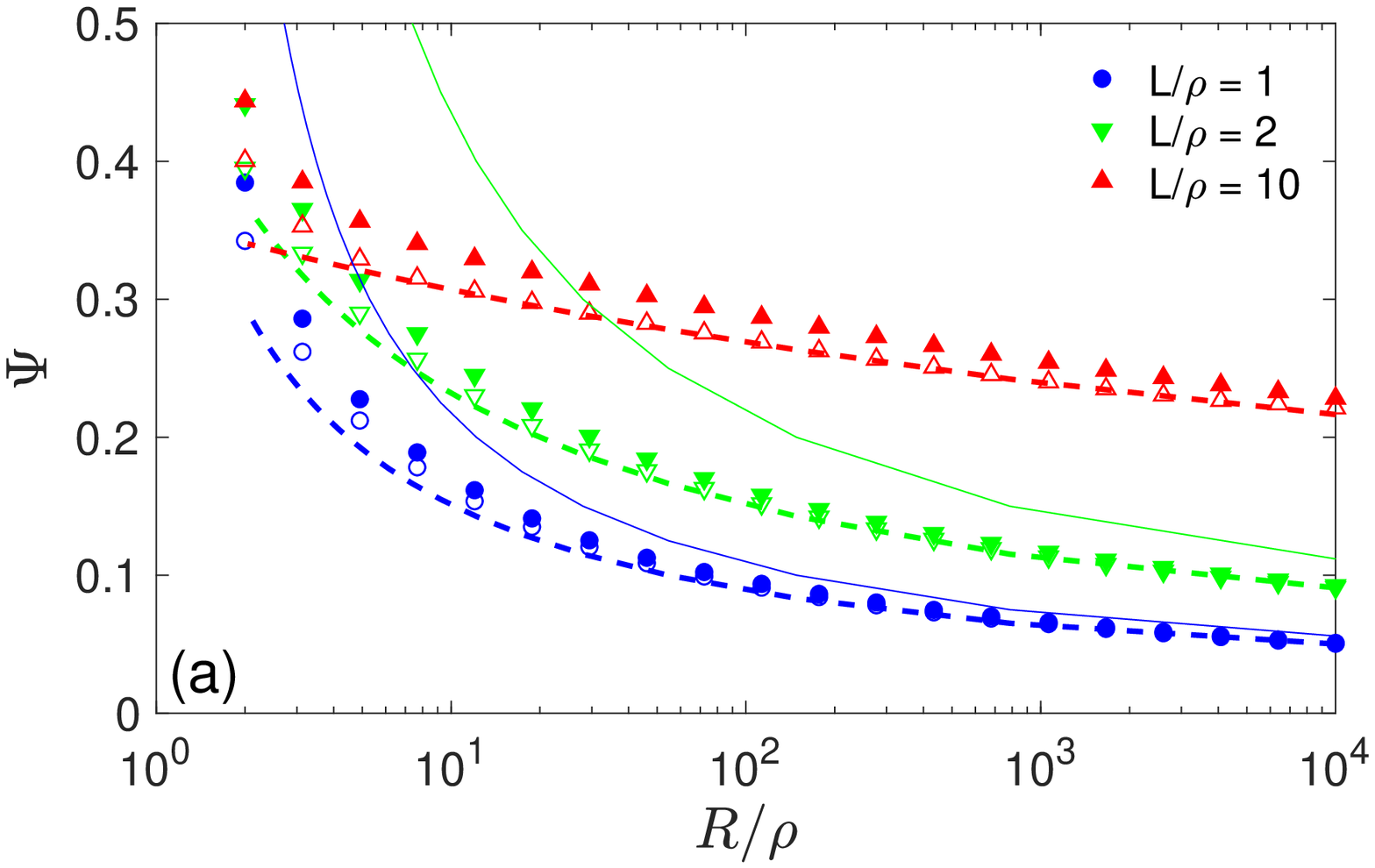}  
\includegraphics[width=85mm]{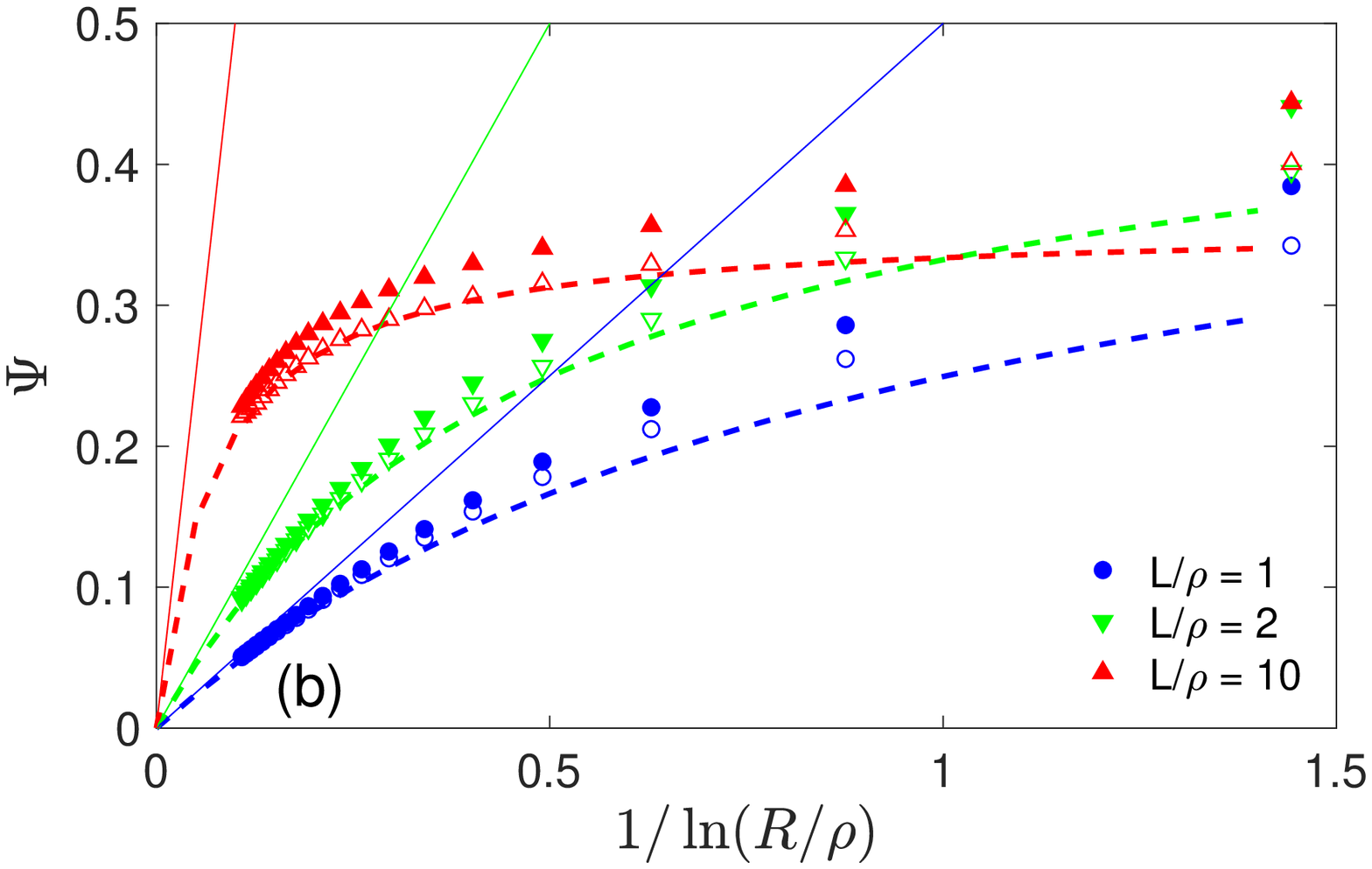}  
\end{center}
\caption{
{\bf (a)} The normalized flux $\Psi$ as a function of $R/\rho$, and
three values of $L/\rho$.  Full symbols show the exact semi-analytical
solution in Eq. (\ref{eq:Psi_exact}) obtained by truncating the
matrices to the size $500 \times 500$ (we also checked that results
are very close for truncation to $1000 \times 1000$); empty symbols
present the self-consistent approximation (\ref{eq:Psi_SCA}), in which
the series is truncated after 10000 terms; thick dashed lines
illustrate the explicit approximation (\ref{eq:Psi_SCA3}); thin solid
lines indicate Eq. (\ref{eq:J0flux}) for two co-axial cylinders (note
that the red curve corresponding to $L/\rho = 10$ lies above $0.5$ and
is thus not visible).  {\bf (b)} The same plot shown as a function of
$1/\ln(R/\rho)$.  }
\label{fig:flux2}
\end{figure}

Unfortunately, the exact semi-analytical solution for $C(r,z,\varphi)$
(Appendix \ref{sec:solution}) is difficult to analyze in the most
relevant regime $R\gg L \gg \rho$.  Even its numerical computation
becomes time consuming because the truncation size of the
infinite-dimensional matrix has to be large.  To get a more suitable
expression for the flux, we apply the self-consistent approximation
(also known as constant-flux approximation) originally devised by
Shoup, Lipari and Szabo \cite{Shoup81} and then extensively adapted to
first-passage time problems
\cite{Grebenkov17a,Grebenkov17,Grebenkov18}.  The accuracy of this
approximation was investigated in \cite{Traytak95}.  The approximation
consists in replacing the mixed Dirichlet-Neumann conditions
(\ref{eq:Eq3}, \ref{eq:Eq4}) on the lateral boundary at $z = 0$ by an
effective inhomogeneous Neumann condition.  The modified boundary
value problem is simpler and admits an exact explicit solution (see
Appendix \ref{sec:SCA} for details).  In particular, we derived the
following expression for the normalized flux $\Psi$:
\begin{equation}  \label{eq:Psi_SCA}
\Psi_{\rm sca} = \frac{\rho}{16R} \left[\sum\limits_{k=0}^\infty \frac{J_1^2(\alpha_{0k}\rho/R)}{\alpha_{0k}^3 J_1^2(\alpha_{0k})} \, 
\biggl(1 + \frac{e^{-\alpha_{0k}L/R}}{\sinh(\alpha_{0k}L/R)}\biggr)\right]^{-1} ,
\end{equation}
where $\alpha_{0k}$ ($k = 0,1,2,\ldots$) are the positive zeros of the
Bessel function $J_0(z)$ of the first kind.  The exact solution of the
modified problem, $\Psi_{\rm sca}$, turns out to be a good
approximation for the factor $\Psi$ of the original problem, as
illustrated by Fig. \ref{fig:flux2} (compare full and empty symbols).
Moreover, this approximation is getting more accurate when $\rho$ is
decreased (or $R$ and $L$ are increased), i.e., in the most relevant
regime.

Most importantly, the self-consistent approximation provides
theoretical insights onto the macroscopic reaction rate.  In
particular, the monotonous decrease of the function $e^{-z}/\sinh z$
in Eq. (\ref{eq:Psi_SCA}) implies a monotonous increase of $\Psi_{\rm
sca}$ and its approach to a constant as $L$ increases.  In other
words, a larger separation between lateral boundaries increases the
diffusive flux onto the reactive region.  Moreover, the two sums in
Eq. (\ref{eq:Psi_SCA}) can be approximately evaluated in the regime $R
\gg L \gg \rho$.  In fact, the first, slowly converging sum in
Eq. (\ref{eq:Psi_SCA}) can be accurately approximated for $\rho/R
\lesssim 0.1$ as
\begin{equation}
\sum\limits_{k=0}^\infty \frac{J_1^2(\alpha_{0k}\rho/R)}{\alpha_{0k}^3 J_1^2(\alpha_{0k})} \simeq \frac{2\rho}{3\pi R} + O\bigl((\rho/R)^2\bigr)\,.
\end{equation}
In turn, the second sum is exponentially converging so that for small
$\rho/R$, it can be approximated as
\begin{equation*}
\frac{\rho^2}{4R^2} \sum\limits_{k=0}^\infty \frac{1}{\alpha_{0k} J_1^2(\alpha_{0k})} \, 
 \frac{e^{-\alpha_{0k}L/R}}{\sinh(\alpha_{0k}L/R)}  
\approx \frac{\rho^2}{8RL} \ln \biggl(\frac{R}{2L}\biggr) \, ,
\end{equation*}
where the last approximation is obtained numerically for small $L/R$.
Combining these asymptotic relations, one gets an approximation
\begin{equation}  \label{eq:Psi_SCA3}
\Psi_{\rm app} \simeq \frac12 \biggl(\frac{16}{3\pi} + \frac{\rho}{L} \ln \biggl(\frac{R}{2L}\biggr) \biggr)^{-1}  
\quad (R \gg L \gg \rho).
\end{equation}
As illustrated in Fig. \ref{fig:flux2}, this approximation is less
accurate than Eq. (\ref{eq:Psi_SCA}) but it is getting more and more
accurate as $R$ increases.  In contrast to the relation
(\ref{eq:Psi_cyl}) for co-axial cylinders, this explicit approximation
captures correctly the dependence of the normalized flux $\Psi$ on
both geometric parameters, $L/\rho$ and $R/\rho$, and can be used to
estimate $\Psi$ without resorting to numerical solutions.  Note also
that the approximation (\ref{eq:Psi_SCA3}) and the expression
(\ref{eq:Psi_cyl}) turn out to be the lower and the upper bounds for
the factor $\Psi$, respectively:
\begin{equation}
\Psi_{\rm app} \leq \Psi \leq \Psi_{\rm cyl}  .
\end{equation}
Although we have no rigorous proof for these inequalities, they can be
used for a rough estimate of the normalized flux $\Psi$.

The approximation (\ref{eq:Psi_SCA3}) highlights the main features of
the reaction rate $J$ in our geometric setting.  In the limit
$L\to\infty$, the factor $\Psi_{\rm sca}$ approaches a constant
$3\pi/32 \approx 0.2945$ which is close to the exact value in
Eq. (\ref{eq:Psi_disk}).  In turn, in the limit $R\to\infty$, the
behavior of the factor $\Psi_{\rm sca}$ becomes similar to
Eq. (\ref{eq:Psi_cyl}) for co-axial cylinders, except that the radius
$\rho$ of the reactive region is replaced by $2L$ under the logarithm.
Most importantly, the approach to the latter limit is extremely slow:
the constant $A = 16/(3\pi) \approx 1.7$ can be neglected when $R/L
\gg 2 \exp(A^{L/\rho})$.  For instance, if $L/\rho = 5$, $R/L$ needs
to be much larger than $10^{6}$.  In other words, whenever $L/\rho
\gtrsim 5$, the co-axial approximation is not applicable, whereas the
approximation (\ref{eq:Psi_SCA3}) yields rather accurate results.  We
emphasize that the limits $L\to\infty$ and $R\to\infty$ cannot be
interchanged, their order is important.

\begin{figure}
\begin{center}
\includegraphics[width=85mm]{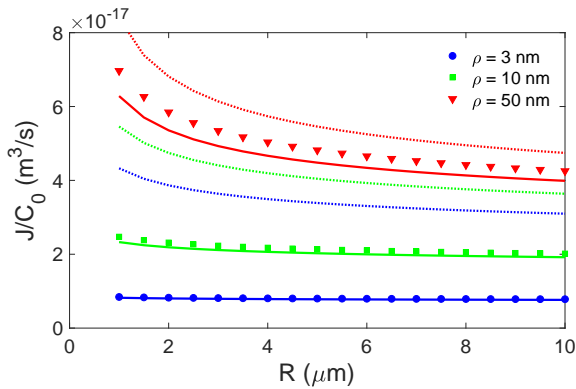} 
\end{center}
\caption{
The diffusive flux $J$ toward the reactive region, normalized by the
concentration $C_0$ at the outer surface, as a function of the outer
radius $R$, with $L = 50$~nm, $D = 800~\mu$m$^2$/s, and three values
of $\rho$: $3$~nm (circles), $10$~nm (squares), and $50$~nm
(triangles).  Solid lines show our approximation (\ref{eq:Psi_SCA3}),
whereas dotted lines indicate Eq. (\ref{eq:J0flux}) for two co-axial
cylinders. }
\label{fig:flux_exper}
\end{figure}

For illustrative purposes, we compute the diffusive flux $J$ for a
realistic set of the model parameters: $D = 800~\mu$m$^2$/s, $L =
50$~nm, $\rho = 3 - 50$~nm, and $R = 1-10~\mu$m.  Figure
\ref{fig:flux_exper} shows the diffusive flux as a function of the
outer radius $R$ for three values of the radius $\rho$: $3$~nm (a
single receptor), $10$~nm (a small group of receptors), and $50$~nm (a
large cluster of receptors).  In the former case, the flux does not
almost depend on the outer radius, as expected for the regime $R \gg L
\gg \rho$.  In turn, when the inner radius $\rho$ becomes comparable
to the inter-cell distance $L$, a weak dependence on $R$ emerges.  One
can see that our approximation (\ref{eq:Psi_SCA3}) accurately captures
this behavior.  For comparison, we also plot the flux from
Eq. (\ref{eq:J0flux}) for two co-axial cylinders.  Although this
formula reproduces qualitatively the behavior of the flux for $\rho$
comparable to $L$, it strongly over-estimates the flux for small
$\rho$.

\section{Discussion}
\label{sec:discussion}

For a particular geometric shape of a junction
(Fig. \ref{fig:scheme}), we obtained the exact semi-analytical
solution for the steady-state concentration of particles, diffusing
from the outer boundary of the junction to a reactive region on the
surface.  We focused on the macroscopic reaction rate $J$ and studied
its dependence on three geometric parameters: the radius $\rho$ of the
reactive region, and the radius $R$ and height $L$ of the junction.
This geometric shape falls in between two commonly studied limits:
unrestricted diffusion in three-dimensional space ($L,R\to
\infty$) and two-dimensional diffusion ($L\to 0$).  We showed that
none of the conventional expressions for the reaction rate is
applicable in the intermediate regime $R\gg L \gg \rho$, which is the
most relevant for many biological applications.  Using the
self-consistent approximation, we managed to derive an explicit simple
approximation for the reaction rate $J$.  For a fixed $L\gg \rho$,
this approximation highlights the extremely slow decay of the reaction
rate with the junction radius $R$.  On one hand, the very weak
dependence on $R$ suggests that this parameter is irrelevant.  On the
other hand, one cannot fully get rid of this parameter by setting
$R\to \infty$ (as in the conventional three-dimensional case) because
the reaction rate would vanish.

The derivation of both semi-analytical solution and self-consistent
approximation relied on modeling the target protein by a reactive disk
located at the center of the lateral surface.  While this assumption
may look oversimplistic, one can argue that the shape and location of
the reactive region are not relevant in the regime $R \gg L \gg
\rho$.  For instance, the normalized flux $\Psi$ changes from $1/\pi
\approx 0.32$ for a disk to $0.5$ for a half-sphere in the half-space
($L,R\to\infty$).  Even if $L$ is not infinitely large, the dependence
on the shape is expected to be weak.  Similarly, the displacement of
the reactive region from the center of the lateral surface would
change the distance $R-\rho$ from the source to this region.  If this
distance is still much larger than $L$ and $\rho$, the reaction rate
should not be much affected (see the related discussion in
Ref. \cite{Gill2011} for planar diffusion, for which an explicit
solution can be obtained by a conformal mapping even for a
non-concentric reactive region; see also \cite{Grebenkov16} and
references therein for other diffusion-reaction applications of
conformal mapping).

Another simplification consisted in considering perfect reactions
occuring immediately upon the first encounter with the reactive
region.  In practice, a particle arriving onto the reactive region has
to overcome an energetic barrier to form a complex, and may fail to
react and thus resume its diffusive motion.  This mechanism of
imperfect reactions can be accounted for via a finite reactivity
$\kappa$ in the Robin boundary condition on the reactive region
\cite{Collins49,Sano79,Sapoval94,Grebenkov06,Grebenkov07,Singer08,Bressloff08}. 
For instance, the Smoluchowski reaction rate $J_S$ is reduced by the
factor $1 + D/(\kappa\rho)$ for imperfect reactions \cite{Collins49}.
Both the semi-analytical solution and self-consistent approximation
can be easily adapted to this problem (see Appendices
\ref{sec:solution} and \ref{sec:SCA} for details).  In particular,
applying the same analysis to the self-consistent approximation
(\ref{eq:Psi_sca_kappa}), one can extend the approximation
(\ref{eq:Psi_SCA3}) to imperfect reactions
\begin{equation}  \label{eq:Psi_SCA3_kappa}
\Psi_{\rm app} \simeq \frac12 \left[\frac{2D}{\kappa\rho} + 
\frac{16}{3\pi} + \frac{\rho}{L} \ln \biggl(\frac{R}{2L}\biggr) \right]^{-1}  
\quad (R \gg L \gg \rho).
\end{equation}
As for a small sphere in $\R^3$, the binding process with the
disk-like reactive region becomes reaction-limited as $\rho\to 0$ or
$\kappa\to 0$, with the dominant term $2D/(\kappa\rho)$, independently
of other geometric parameters.  Even if the reaction mechanism is
relatively fast so that the term $2D/(\kappa \rho)$ is of order $1$,
its presence can significantly affect the reaction rate.  This
highlights the importance of accounting for imperfect reaction
mechanisms and potential pitfalls of considering reactions as perfect.
Note that we have put forward perfect reactions for clarity of
presentation, bearing in mind the provided extensions for the
imperfect case.

Finally, we focused on a single reactive site located on the lateral
surface.  In biological applications, there are typically many
reactive sites distributed over the surface.  Even if these sites are
well separated from each other, the total diffusive flux is not equal
to the flux to a single site multiplied by the number of sites.  In
fact, reactive sites compete for capturing the diffusing particles
that yields long-range diffusive interactions between reactive sites
\cite{Traytak92,Galanti16,Traytak18}.  Moreover, the reactive
region located closer to the outer source can (partly) screen the
reactive sites in the central region
\cite{Sapoval94,Felici03,Gill2011}.  As a consequence, the analysis of
the steady-state diffusion equation with multiple reactive regions is
much more involved and often relies on numerical solutions.  Since
this problem is beyond the scope of the paper, we only mention that a
region of the lateral surface covered by uniformly distributed
reactive sites can be modeled as a large partially reaction region.
In this homogenized problem, the partial reactivity $\kappa$ accounts
for eventual reflections of diffusing particles on passive regions of
the lateral surface.

We have solved explicitly the diffusion to capture problem within a
sheet.  This problem is particularly relevant to the diffusion of
signaling molecules within organelles and to that of glutamate and
other neurotransmitters in the brain extracellular spaces.  The
problem is solved semi-analytical as a function of reactive region
size, sheet thickness, and radial distance to the source. Besides the
exact solutions, simple expressions for the thin- and thick-sheet
asymptotics as well as the large radial distance-to-source are
provided. Some biologically-relevant conclusions are obtained. In
particular, even though the flux is monotonically decreasing with the
radial distance-to-source, the flux decay is very slow, making the
reaction rate practically insensitive to this distance.  Thus, it is
possible for a cell system to dramatically reduce the area devoid of
reactive sites without a substantial effect on the adsorption flux.
Another important result is shown for the effect of the sheet
thickness.  For very thin sheets, when the thickness is smaller than
the radius of the reactive region, the flux is proportional to the
sheet thickness and practically insensitive to that radius.  Thus in
this system, the flux to capture to a single protein is in essence the
same as that of capture to a cluster of molecules and, as a
consequence, the cell would not gain anything by increasing the number
of target proteins within a single site. On the other hand, for large
sheet thicknesses, the flux is proportional to the radius of the
reactive region.  Finally, the solution to the case where both the
diffusion and adsorption rate play roles, i.e., for the imperfect
reaction is also provided.  This case is particularly relevant for
glutamate transporters that are known to be limited by the velocity of
glutamate transport across the astrocytic membrane.

\begin{acknowledgments}
DG acknowledges the support under Grant No. ANR-13-JSV5-0006-01 of the
French National Research Agency.
\end{acknowledgments}


\setcounter{section}{0}
\renewcommand{\thesection}{\Alph{section}}
\renewcommand{\thesubsection}{\arabic{subsection}}
\section*{APPENDICES}
\section{Semi-analytical solution}
\label{sec:solution}

In this Appendix, we provide the mathematical derivation of the
macroscopic reaction rate on the target protein (the reactive region
$\Gamma$), which is located at the center of the surface, i.e., it is
concentric with the junction domain $\Omega$: $\Gamma = \{ 0 < r <
\rho, ~ z = 0\}$. 
As a consequence, one can drop the dependence on the angular
coordinate $\varphi$, and the boundary value problem (\ref{eq:Eq})
becomes
\begin{subequations}  \label{eq:eq}
\begin{eqnarray}
\label{eq:eq1}
\biggl(\partial_r^2 + \frac{1}{r} \partial_r + \partial_z^2\biggr) u(r,z) &=& 0, \\
\label{eq:eq2}
u_{|r=R} &=& 1, \\
\label{eq:eq3}
u_{|z=0} &=& 0 \quad (0 < r<\rho), \\
\label{eq:eq4}
(\partial_z u)_{|z=0} &=& 0 \quad (\rho < r < R),\\
\label{eq:eq5}
(\partial_z u)_{|z=L} &=& 0 ,
\end{eqnarray}
\end{subequations}
from which $C(r,z,\varphi) = C_0 u(r,z)$.  While the problem is
classic, the mixed Dirichlet-Neumann boundary conditions
(\ref{eq:eq3}, \ref{eq:eq4}) present the major technical difficulty.

To overcome this difficulty, it is convenient to split the domain
$\Omega$ into two parts: $\Omega_1 = \{ 0 < r < \rho, ~ 0 < z < L\}$
and $\Omega_2 = \{ \rho < r < R, ~ 0 < z < L\}$.  In the inner part
$\Omega_1$, we search a solution in the form satisfying
Eqs. (\ref{eq:eq1}, \ref{eq:eq3}, \ref{eq:eq5}):
\begin{equation}  \label{eq:u1}
u_1(r,z) = \sum\limits_{n=1}^\infty c_n^{(1)} \, v_n^{(1)}(r) \sin(\alpha_n^{(1)} z/L)  ,
\end{equation}
with $\alpha_n^{(1)} = \pi(n-1/2)$,
\begin{equation}
v_n^{(1)}(r) = \frac{I_0(\alpha_n^{(1)} r/L)}{I_0(\alpha_n^{(1)} \rho/L)} \,,
\end{equation}
$c_n^{(1)}$ are unknown coefficients, and $I_\nu(z)$ are the modified
Bessel functions of the first kind.
In the outer part $\Omega_2$, we search a solution in the form
satisfying Eqs. (\ref{eq:eq1}, \ref{eq:eq2}, \ref{eq:eq4},
\ref{eq:eq5}):
\begin{equation}  \label{eq:u2}
u_2(r,z) = 1 + c_0^{(2)} \ln(r/R) + \sum\limits_{n=1}^\infty c_n^{(2)} \, v_n^{(2)}(r) \cos(\alpha_n^{(2)} z/L)  ,
\end{equation}
where $\alpha_n^{(2)} = \pi n$, 
\begin{equation}
v_n^{(2)}(r) = K_0(\alpha_n r/L) - I_0(\alpha_n r/L) \frac{K_0(\alpha_n R/L)}{I_0(\alpha_n R/L)} \,,
\end{equation}
$c_n^{(2)}$ are unknown coefficients, and $K_\nu(z)$ are the modified
Bessel functions of the second kind.

\begin{figure}
\begin{center}
\includegraphics[width=85mm]{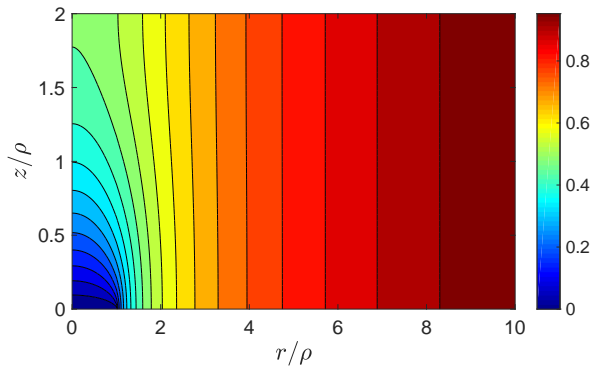} 
\end{center}
\caption{
Concentration $u(r,z)$ for $\rho = 1$, $R = 10$, and $L = 2$.}
\label{fig:concentration}
\end{figure}

The unknown coefficients can be determined by matching two solutions
at $r = \rho$:
\begin{subequations}
\begin{eqnarray}  
\label{eq:matching1}
u_1(\rho,z) &=& u_2(\rho,z),  \\ 
\label{eq:matching2}
(\partial_r u_1(r,z))_{|r=\rho} &=& (\partial_r u_2(r,z))_{|r=\rho} \,.
\end{eqnarray}
\end{subequations}
Substituting $u_1$ and $u_2$ into the first relation, multiplying by
$\sin(\alpha_m^{(1)} z/L)$ and integrating over $z$ from $0$ to $L$,
one gets an infinite system of linear equations,
\begin{equation}
\frac{2}{\alpha_m^{(1)}} \bigl(1 + c_0^{(2)}\ln (\rho/R)\bigr) + \sum\limits_{n=1}^\infty c_n^{(2)} v_n^{(2)}(\rho) B_{nm} = c_m^{(1)} 
\end{equation}
for each $m = 1,2,\ldots$, where we used that $v_m^{(1)}(\rho) = 1$,
and $B$ is the infinite-dimensional matrix with elements
\begin{eqnarray}  \nonumber
B_{nm} &=& \frac{2}{L} \int\limits_0^L dz \, \sin(\alpha_m^{(1)} z/L) \cos(\alpha_n^{(2)} z/L)   \\
\label{eq:B}
&=& \frac{2\alpha_m^{(1)}}{[\alpha_m^{(1)}]^2 - [\alpha_n^{(2)}]^2} \,, 
\end{eqnarray}
because $\cos\alpha_n^{(1)} = \sin \alpha_n^{(2)} = 0$.  Note that
\begin{equation}
BB^\dagger = I , \qquad B^\dagger B = I + C,
\end{equation}
where $I$ is the identity matrix, $\dagger$ denotes the matrix
transposition, and
\begin{equation} \label{eq:Cmatrix}
C_{mn} = - \frac{2}{\alpha_m^{(1)} \alpha_n^{(1)}} \,.
\end{equation}
Next, substituting $u_1$ and $u_2$ into Eq. (\ref{eq:matching2}),
multiplying by $\cos(\alpha_m^{(2)} z/L)$ and integrating over $z$
from $0$ to $L$, one gets another infinite system of linear equations,
\begin{equation}  \label{eq:auxil1}
c_m^{(2)} \bigl(\partial_r v_m^{(2)}(r)\bigr)_{|r=\rho} = \sum\limits_{n=1}^\infty c_n^{(1)} \bigl(\partial_r v_n^{(1)}(r)\bigr)_{|r=\rho} B_{mn}
\end{equation}
with $m = 1,2,\ldots$.  Finally, the integral of
Eq. (\ref{eq:matching2}) over $z$ from $0$ to $L$ yields
\begin{equation}  \label{eq:c0}
c_0^{(2)} = \rho \sum\limits_{n=1}^\infty \frac{c_n^{(1)} (\partial_r v_n^{(1)}(r))_{|r=\rho}}{\alpha_n^{(1)}} \,.
\end{equation}
Combining these equations, one gets a closed infinite system of linear
equations for unknowns $c_n^{(1)}$:
\begin{eqnarray*}
&& c_m^{(1)} = \frac{2}{\alpha_m^{(1)}} + \frac{2\rho \ln(\rho/R)}{\alpha_m^{(1)}} \sum\limits_{n=1}^\infty 
c_n^{(1)} \frac{(\partial_r v_n^{(1)}(r))_{|r=\rho}}{\alpha_n^{(1)}} \\
&+& \sum\limits_{n=1}^\infty c_n^{(1)} (\partial_r v_n^{(1)}(r))_{|r=\rho} \sum\limits_{n'=1}^\infty 
B_{n'n} \frac{v_{n'}^{(2)}(\rho)}{(\partial_r v_{n'}^{(2)}(r))_{|r=\rho}} B_{n'm} .
\end{eqnarray*}
Introducing diagonal matrices  
\begin{eqnarray}
V_{mn}^{(1)} &=& \delta_{mn}\, \frac{1}{L \,(\partial_r v^{(1)}_n(r))_{|r=\rho}}  \,,\\
V_{mn}^{(2)} &=& \delta_{mn}\, \frac{v_n^{(2)}(\rho)}{L (\partial_r v_n^{(2)}(r))_{|r=\rho}} \,,  
\end{eqnarray}
one can rewrite the above equations in a matrix form as
\begin{equation}
{\bf c}^{(1)} = 2{\bf b} + (\eta C + B^\dagger V^{(2)} B) (V^{(1)})^{-1} {\bf c}^{(1)} ,
\end{equation}
where
\begin{equation}  \label{eq:eta}
{\bf b}_n = \frac{1}{\alpha_n^{(1)}} \,,  \qquad
\eta = (\rho/L) \ln(R/\rho) ,
\end{equation}
and the matrix $C$ is defined by Eq.~(\ref{eq:Cmatrix}).  One gets
thus
\begin{equation}  \label{eq:cm1}
c_m^{(1)} = 2 \bigl[V^{(1)} X {\bf b} \bigr]_m ,
\end{equation}
where
\begin{equation} \label{eq:X}
X = \bigl(V^{(1)} - (\eta C + B^\dagger V^{(2)} B) \bigr)^{-1} .
\end{equation}
From Eq. (\ref{eq:auxil1}), one also gets
\begin{equation}
c_m^{(2)} = \frac{2}{L (\partial_r v_m^{(2)}(r))_{r=\rho}}  \bigl[B X {\bf b} \bigr]_m ,
\end{equation}
while $c_0^{(2)}$ is given according to Eq. (\ref{eq:c0}) as
\begin{equation}
c_0^{(2)} = \frac{2\rho}{L} \bigl( {\bf b} \cdot X {\bf b} \bigr).
\end{equation}
The found coefficients $c_n^{(1)}$ and $c_n^{(2)}$ fully determine the
solution $u(r,z)$ of the boundary value problem (\ref{eq:eq}).  Figure
\ref{fig:concentration} shows the concentration profile $u(r,z)$ for a
reactive disk of radius $\rho = 1$.  One can see how the concentration
drops from $1$ at the outer cylinder (at $r = R = 10$) to $0$ at the
reactive region.

The diffusive flux onto the reactive region reads
\begin{eqnarray*}
J &=& 2\pi C_0 D \int\limits_0^{\rho} dr \,r \, (\partial_z u_1(r,z))_{z=0} \\
&=& 2\pi C_0 D \rho \sum\limits_{n=1}^\infty c_n^{(1)} \frac{I_1(\alpha_n^{(1)}\rho/L)}{I_0(\alpha_n^{(1)}\rho/L)} \\
&=& 2\pi C_0 D \rho \sum\limits_{n=1}^\infty c_n^{(1)} \frac{L(\partial_r v_n^{(1)}(r))_{r=\rho}}{\alpha_n^{(1)}} \\
&=& 4\pi C_0 D \rho \bigl({\bf b} \cdot X {\bf b} \bigr).
\end{eqnarray*}
Dividing this expression by the Smoluchowski flux in
Eq. (\ref{eq:Smoluchowski}), one finds the normalized flux $\Psi$
accounting for the shape of the domain:
\begin{equation}  \label{eq:Psi_exact}
\Psi = \bigl({\bf b} \cdot X {\bf b} \bigr).
\end{equation}
By construction, $\Psi$ is equal to $1$ for a reactive sphere of
radius $\rho$ in the three-dimensional space.  In our setting, the
normalized flux is a function of two geometric parameters, $R/\rho$
and $L/\rho$.

While the above solution of the boundary value problem (\ref{eq:eq})
and the consequent expression (\ref{eq:Psi_exact}) for the normalized
flux $\Psi$ are exact, they involve the inversion of an
infinite-dimensional matrix in Eq. (\ref{eq:X}) that requires
numerics.  For this reason, the presented solution is called
semi-analytical.  In practice, one needs to truncate
infinite-dimensional matrices and vectors and then to compute
numerically the normalized flux $\Psi$ and thus the flux $J$.  The
accuracy of this computation is controlled by the truncation size and
can be easily verified.  Although the dependence on the geometric
parameters $R/\rho$ and $L/\rho$ is ``hidden'' by the matrix
inversion, this semi-analytical solution is easily computable and thus
allows one to explore the shape dependence.

\subsection{Limit $R \to\infty$}

In the limit $R\to \infty$, one has
\begin{equation}
v_n^{(2)}(r) = K_0(\alpha_n^{(2)} r/L) ,
\end{equation}
so that
\begin{eqnarray}
V_{mn}^{(1)} &=& \delta_{mn} \, \frac{I_0(\alpha_n^{(1)} \rho/L)}{\alpha_n^{(1)} \, I_1(\alpha_n^{(1)} \rho/L)} \,, \\
V_{mn}^{(2)} &=& - \delta_{mn} \, \frac{K_0(\alpha_n^{(2)} \rho/L)}{\alpha_n^{(2)} \, K_1(\alpha_n^{(2)} \rho/L)} \,.
\end{eqnarray}
While all the matrices remain well defined in this limit, the factor
$\eta$ in the matrix $X$ in Eq. (\ref{eq:X}) diverges logarithmically.
Qualitatively, one can thus expect that the matrix $X$ vanishes
logarithmically as well.

To clarify this point, we consider the regime $L \ll \rho$, for which
\begin{eqnarray*}
V_{mn}^{(1)} &\longrightarrow& (V_0^{(1)})_{mn} = \frac{\delta_{mn}}{\alpha_n^{(1)}} \, , \\
V_{mn}^{(2)} &\longrightarrow& (V_0^{(2)})_{mn} = - \frac{\delta_{mn}}{\alpha_n^{(2)}} 
\end{eqnarray*}
(this is also true in the limit $n\to \infty$).

In a first attempt, one can try to neglect all matrices in the
expression (\ref{eq:X}) for $X$, except for the dominant term $-\eta
C$.  However, such an approximation is useless as the matrix $C$ is
not invertible.  For this reason, we neglect only the matrix
$\B^\dagger V^{(2)} B$.  In fact, one can show that this matrix is
asymptotically comparable to the matrix $C$ and thus can be neglected
as compared to $C$ due to the large factor $\eta$ in front of $C$.
The remaining matrix $V^{(1)} - \eta C$, truncated to the size
$N\times N$, can be inverted explicitly, i.e.,
\begin{equation}
\tilde{X}_{mn}^{(N)} = \biggl[\bigl( V^{(1)} - \eta C\bigr)^{-1}\biggr]_{mn} = \delta_{mn} \alpha_n^{(1)} - \frac{2\eta}{1 - 2\eta A_N},
\end{equation}
where $A_N = 1/\alpha_1^{(1)} + \ldots + 1/\alpha_N^{(1)}$ (the sign
tilde highlights that this is an approximation, in which the matrix
$\B^\dagger V^{(2)} B$ was neglected).  As a consequence, the
normalized flux $\Psi$ becomes
\begin{equation}
\Psi_N = \frac{A_N}{1 + 2\eta A_N} \,.
\end{equation}
In the limit $N\to\infty$, the series of $1/\alpha_k^{(1)}$ diverges
logarithmically, i.e., $A_N \to \infty$, from which we retrieve the
normalized flux $\Psi$ for two co-axial cylinders:
\begin{equation}
\Psi_N \to \Psi_{\rm cyl} = \frac{1}{2\eta} = \frac{L/\rho}{2\ln(R/\rho)} \,.
\end{equation}
This is expected because the geometric setting $R \gg \rho \gg L$
resembles to two-dimensional diffusion.  However, getting corrections
to this limit due to the matrix $\B^\dagger V^{(2)} B$ is a difficult,
technically involved problem which is beyond the scope of the paper.

\subsection{Limit $R \to \rho$}
\label{sec:Rrho}

When $R\to \rho$, the outer subdomain $\Omega_2$ shrinks.  Setting $R
= \rho(1 + \ve)$, the first-order expansions in powers of $\ve$ are
\begin{equation}
\eta \simeq \rho \ve/L, \qquad V_{mn}^{(2)} = - \delta_{mn} \rho \ve/L ,
\end{equation}
so that
\begin{equation}
X = \bigl( V^{(1)} + (\rho \ve/L) I\bigr)^{-1} .
\end{equation}
As a consequence, Eq. (\ref{eq:cm1}) implies that $c_n^{(1)} \to
2/\alpha_n^{(1)}$ as $\ve \to 0$ and thus
\begin{equation}  \label{eq:u_Rrho}
u(r,z) = u_1(r,z) = 2\sum\limits_{n=1}^\infty \frac{v_n^{(1)}(r)}{\alpha_n^{(1)}} \, \sin(\alpha_n^{(1)} z/L) \,.
\end{equation}
Moreover, Eq. (\ref{eq:Psi_exact}) yields the diffusive flux and the
factor $\Psi$:
\begin{equation}
\Psi = \sum\limits_n  \frac{1}{(\alpha_n^{(1)})^2} \, \biggl( \frac{\rho \ve}{L} + 
\frac{I_0(\alpha_n^{(1)}\rho/L)}{\alpha_n^{(1)} I_1(\alpha_n^{(1)}\rho/L)} \biggr)^{-1} .
\end{equation}
This relation shows that the diffusive flux diverges logarithmically
with $\ve$.  We checked numerically that 
\begin{equation}
\Psi \simeq - \frac{1}{\pi} \ln \ve + O(1)  \qquad (\ve \to 0), 
\end{equation}
with $\ve = (R - \rho)/\rho$, and the constant term depending on
$\rho/L$.  The divergence of the flux also follows from a direct
computation of $J$ from the exact solution in Eq. (\ref{eq:u_Rrho}).
In mathematical terms, the divergence is related to the fact that at
the circle $\{r = \rho,~ z = 0\}$, the boundary conditions $C = C_0$
at $r = \rho$ and $C = 0$ at $z = 0$ contradict to each other.

\subsection{Extension to partial reactivity}
\label{eq:partial}

When the target protein is partially reactive (i.e., when the reaction
does not occur immediately upon the arrival onto the reactive region),
the Dirichlet boundary condition (\ref{eq:eq3}) should be replaced by
a more general Robin boundary condition
\begin{equation}  \label{eq:Robin}
\bigl[D (\partial_z u) - \kappa u\bigr]_{|z=0} = 0 \quad (0 < r < \rho),
\end{equation}
where $\kappa$ is the reactivity
\cite{Collins49,Sano79,Sapoval94,Grebenkov06,Grebenkov07,Singer08,Bressloff08},
which can vary from $0$ (no reaction) to infinity (perfect reaction,
as considered in the main text).  This extension affects only the
solution (\ref{eq:u1}) in the inner subdomain, in which
$\sin(\alpha_n^{(1)} z/L)$ is replaced by
$\cos(\alpha_n^{(1)}(L-z)/L)$, where $\alpha_n^{(1)}$ are now obtained
as solutions of the trigonometric equation:
\begin{equation}
\frac{D}{\kappa L} \alpha_n^{(1)} \sin \alpha_n^{(1)} = \cos \alpha_n^{(1)} .
\end{equation}
This change also affects the computation of the matrix $B$ for which
Eq. (\ref{eq:B}) is replaced by
\begin{equation}
B_{nm} = \frac{2\alpha_m^{(1)} \, \sin \alpha_m^{(1)}}{[\alpha_m^{(1)}]^2 - [\alpha_n^{(2)}]^2} \,.
\end{equation}

\section{Disk in the half-space}  
\label{sec:half-space}

In the double limit $L\to\infty$ and $R \to \infty$, the problem is
reduced to the classic problem of electrified disk in the
(half)-space, for which the solution was found by Weber (see
\cite{Sneddon}, p. 64).  In our notations, the solution in the upper
half-space reads
\begin{equation}
u(r,z) = 1 - \frac{2}{\pi}\int\limits_0^\infty \frac{d\mu}{\mu} \sin(\mu\rho) \, e^{-\mu z} J_0(\mu r).
\end{equation}
This function satisfies the Laplace equation with the mixed boundary
conditions on the plane $z = 0$:
\begin{equation}
\begin{split}
u(r,z)|_{z=0} & = 0 \quad (0\leq r < \rho),  \\
(\partial_z u(r,z))|_{z=0} & = 0 \quad (r > \rho), \\
\end{split}
\end{equation}
and $u(r,z) \to 1$ as $r\to\infty$ or $z\to\infty$.
The flux density is then
\begin{eqnarray}  \nonumber
j(r) &=& D C_0 \frac{2}{\pi} \int\limits_0^\infty d\mu \sin(\mu\rho) \, J_0(\mu r) \\
&=& \frac{2 D C_0}{\pi \sqrt{\rho^2 - r^2}}  \qquad  (0 \leq r < \rho),
\end{eqnarray}
whereas the diffusive flux is obtained by integration over the disk:
\begin{equation}  \label{eq:Jdisk}
J_{\rm disk} = 4 D C_0 \rho .  
\end{equation}
We retrieved thus the particular case of the Hill formula
\cite{Hill75}.  The flat shape of the disk is reduced the factor
$\Psi$ from $1/2$ for a half-sphere to $\Psi_{\rm disk} = 1/\pi
\approx 0.3183$.

\section{Solution for an oblate spheroid}
\label{sec:oblate}

\begin{figure}
\begin{center}
\includegraphics[width=45mm]{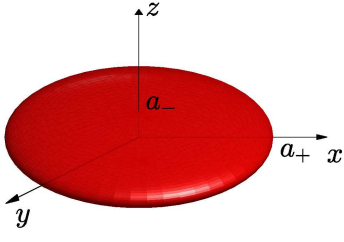}  
\end{center}
\caption{
Oblate spheroid with major and minor semiaxes $a_+$ and $a_-$.}
\label{fig:spheroid}
\end{figure}

As we mentioned in Appendix \ref{sec:half-space}, the particular
choice of the reactive region as a disk reduces the factor $\Psi$.  In
order to illustrate the dependence on the shape of the reactive
region, we recall the solution of the steady-state diffusion equation
for an oblate spheroid in $\R^3$ (see Fig. \ref{fig:spheroid}).  It is
natural to use the oblate spheroidal coordinates
$(\mu,\theta,\varphi)$ which are related to the Cartesian coordinates
via
\begin{equation}
\begin{split}
x & = a \cosh \mu \, \cos \theta \, \cos \varphi , \\
y & = a \cosh \mu \, \cos \theta \, \sin \varphi ,\\
z & = a \sinh \mu \, \sin \theta ,  \\
\end{split}
\end{equation}
whereas
\begin{equation}
\begin{split}
\cosh\mu & = \frac{d_+ + d_-}{2a} \,, \\
\cos \theta & = \frac{d_+ - d_-}{2a} \,,\\
\tan \varphi & = y/x \,, \\
\end{split}
\end{equation}
with $d_{\pm} = \sqrt{(\rho \pm a)^2 + z^2}$, $\rho = \sqrt{x^2 +
y^2}$.  Here $\mu$ varies from $0$ to infinity, $\theta$ from $-\pi/2$ to
$\pi/2$, and $\varphi$ from $0$ to $2\pi$.  Note that the oblate
spheroid at a constant $\mu_0$ is obtained by rotating ellipses about
the $z$-axis.  An ellipse in the $x-z$ plane has a major semiaxis of
length $a\cosh\mu_0$ along the $x$-axis, whereas its minor semiaxis
has length $a \sinh\mu_0$ along the $z$-axis.  The foci of all the
ellipses in the $x-z$ plane are located on the $x$-axis at $\pm a$.
In other words, if the major and minor semiaxes are denoted as
$a_\pm$, one has $a = \sqrt{a_+^2 - a_-^2}$ and $\mu_0 =
\tanh^{-1}(a_-/a_+)$.

Harmonic functions can be decomposed onto the following regular and
singular functions
\begin{equation*}
\begin{cases} P_n^m(i\sinh \mu) \, P_n^m(\sin \theta) e^{im\phi}  \qquad (\textrm{regular}), \cr
Q_n^m(i\sinh \mu) \, P_n^m(\sin \theta) e^{im\phi}   \qquad (\textrm{singular}), \end{cases} 
\end{equation*}
where $P_n^m$ and $Q_n^m$ are the Legendre functions of the first and
second kind, respectively.  Since we are interested in a solution
outside the oblate spheroid, we use only singular functions.  Due to
the symmetry of the boundary conditions, there is only the
contribution from $m = n = 0$, i.e.,
\begin{equation}
u = 1 - \frac{Q_0(i \sinh\mu)}{Q_0(i \sinh \mu_0)} \,,
\end{equation}
where $\mu_0$ determines the boundary $\partial\Omega$ of the oblate
spheroid, and $Q_0(z) = \frac12 \ln \bigl(\frac{z+1}{z-1}\bigr)$.
This function solves the following problem
\begin{equation}
\begin{split}
\Delta u & = 0 \quad (\mu > \mu_0), \\
u|_{\partial\Omega} & = 0  \quad (\mu = \mu_0), \\
u|_{\|x\|\to\infty} & \to 1 . \\
\end{split}
\end{equation}
Note that
\begin{equation}
Q_0(iz) = - \frac{i}{2} \cos^{-1}\biggl(\frac{z^2-1}{z^2+1}\biggr),
\end{equation}
so that
\begin{eqnarray}  \label{eq:Q0}
Q_0(i\sinh\mu) &=& -\frac{i}{2} \cos^{-1}\biggl(1 - \frac{2}{\cosh^2\mu}\biggr) \\  \nonumber
&=& -\frac{i}{2} \cos^{-1}\biggl(\frac{(d_+ + d_-)^2 - 8a^2}{(d_+ + d_-)^2}\biggr) .
\end{eqnarray}

The flux density onto the oblate spheroid is then
\begin{eqnarray}  \nonumber
j &=& - C_0 D (\partial_n u)|_{\partial\Omega}= \biggl(\frac{C_0 D}{h_\mu} \partial_\mu u\biggr)|_{\mu=\mu_0} \\
&=& - \frac{C_0 D i}{h_{\mu_0} \, \cosh \mu_0 \, Q_0(i\sinh \mu_0)} \,,
\end{eqnarray}
where $h_\mu = a \sqrt{\sinh^2 \mu + \sin^2 \theta}$.  The diffusive flux
is then
\begin{eqnarray}  
J &=& \int\limits_\Gamma d\theta d\phi \, h_\theta \, h_\phi \, j   \\   \nonumber
&=& 2\pi C_0 D a \int\limits_{-\pi/2}^{\pi/2} d\nu  \frac{i \, \cos \theta}{Q_0(i\sinh \mu_0)}
= \frac{4\pi C_0 D a i}{Q_0(i\sinh\mu_0)} \,,
\end{eqnarray}
where $h_\theta = a \sqrt{\sinh^2 \mu + \sin^2 \theta}$ and $h_\phi = a
\cosh \mu$.  Using the relation $\mu_0 = \tanh^{-1}(a_-/a_+)$ and
Eq. (\ref{eq:Q0}), one gets
\begin{equation}
Q_0(i\sinh\mu_0) = - \frac{i}{2} \cos^{-1}\biggl(\frac{2 a_-^2}{a_+^2} - 1\biggr) = -i \cos^{-1}\bigl(a_-/a_+\bigr).
\end{equation}
One retrieves thus the explicit form of the diffusive flux discussed
in \cite{Berezhkovskii07}
\begin{equation}
J = 4\pi C_0 D  \,\frac{\sqrt{a_+^2 - a_-^2}}{\cos^{-1}(a_-/a_+)} \,.
\end{equation}
In particular, one gets $J = 8 C_0 D a_+$ in the limit $a_- \to 0$, as
expected for the disk of radius $a_+$.  In the limit $a_- \to a_+$,
one can write $a_- = a_+(1-\ve)$ and then $\cos^{-1}(1-\ve) \simeq
2\sqrt{\ve}$, so that the Smoluchowski rate is retrieved.  Dividing
the flux by the Smoluchowski rate for a sphere of radius $a_+$, one
gets
\begin{equation}
\Psi_{\rm oblate} = \frac{\sqrt{1 - (a_-/a_+)^2}}{\cos^{-1}(a_-/a_+)} \,,
\end{equation}
which varies from $2/\pi$ at $a_-/a_+ = 0$ (the disk) to $1$ at
$a_-/a_+ = 1$ (the sphere).  This factor should be halved if one
considers a half of an oblate spheroid in the half-space.

\section{Self-consistent approximation}
\label{sec:SCA}

We also provide an approximate but explicit solution to the problem
based on the self-consistent approximation originally developed by
Shoup, Lipari and Szabo \cite{Shoup81} and then extensively adapted to
first-passage time problems
\cite{Grebenkov17a,Grebenkov17,Grebenkov18}.  Within the
self-consisted approximation, the Dirichlet boundary condition at the
target protein is replaced by an approximate Neumann condition with an
unknown constant flux density $\hat{j}$.  In other words, the mixed
Dirichlet-Neumann boundary conditions (\ref{eq:eq3}, \ref{eq:eq4}) are
replaced by inhomogeneous Neumann condition
\begin{equation}  \label{eq:inhom}
D (\partial_z \hat{u})_{|z = 0} = \hat{j} \, \Theta(\rho - r),
\end{equation}
where $\Theta$ is the Heaviside step function.  The solution of the
modified boundary value problem (\ref{eq:eq1}, \ref{eq:eq2},
\ref{eq:eq5}, \ref{eq:inhom}), denoted as $\hat{u}$, can be expressed
through the Green function $G(\x,\x')$ in the capped cylinder:
\begin{equation}
\hat{u}(\x) = 1 - \frac{\hat{j}}{D} \int\limits_\Gamma d\x' \, G(\x,\x'),
\end{equation}
where $\Gamma$ is the reactive region, and $\x = (r,z,\varphi)$ in
cylindrical coordinates.  The Green function satisfies
\begin{equation}
- \Delta_{\x} G(\x,\x') = \delta(\x-\x') ,
\end{equation}
subject to the Dirichlet boundary condition $G(\x,\x') = 0$ at $r= R$,
and Neumann boundary conditions $\partial_z G(\x,\x') = 0$ at $z = 0$
and $z = L$.  The latter can be expressed via the corresponding
Laplacian eigenfunctions
\begin{equation}
G(\x,\x') = \sum\limits_{m,n,k} \lambda_{mnk}^{-1} u_{mnk}(\x) \, u_{mnk}^*(\x'),
\end{equation}
where 
\begin{eqnarray*}
u_{mnk}(r,z,\varphi) &=& c_{mnk} \, J_n(\alpha_{nk}r/R) \, \cos(\pi m z/L) \, e^{in\varphi} , \\
\lambda_{mnk} &=& \alpha_{nk}^2/R^2 + \pi^2 m^2/L^2 ,
\end{eqnarray*}
with $m = 0,1,\ldots$, $n = 0,1,\ldots$, $k = 0,1,\ldots$, and
$c_{mnk}$ are the normalization constants,
\begin{equation}
c_{mnk} = \frac{\epsilon_m}{\sqrt{\pi L R^2} \, J_{n+1}(\alpha_{nk})}
\end{equation}
(with $\epsilon_m = \sqrt{2}$ for $m >0$, and $\epsilon_0 = 1$), and
$\alpha_{nk}$ are the positive zeros of the Bessel function $J_n(z)$.
We get thus
\begin{widetext}
\begin{eqnarray*}
\hat{u}(r,z) &=& 1 - \frac{\hat{j}}{D} \sum\limits_{m,n,k} \lambda_{mnk}^{-1} u_{mnk}(r,z,\varphi) \int\limits_0^{\rho} dr' \, r' 
\int\limits_0^{2\pi} d\varphi' \, u_{mnk}^*(r',0,\varphi') \\
&=& 1 - \frac{2\pi \hat{j}}{D} \sum\limits_{m,k} \frac{c_{m0k}^2}{\lambda_{m0k}} J_0(\alpha_{0k}r/R) \cos(\pi m z/L) \int\limits_0^{\rho} dr' \, r' 
J_0(\alpha_{0k} r'/R)  \\
&=& 1 - \frac{2\pi \hat{j}}{D} \sum\limits_{m,k} \frac{c_{m0k}^2}{\lambda_{m0k}} J_0(\alpha_{0k}r/R) \cos(\pi m z/L) 
\frac{\rho J_1(\alpha_{0k} \rho/R)}{\alpha_{0k}/R} \\
&=& 1 - \frac{2\hat{j} \rho}{D L R} \sum\limits_{m,k} \frac{\epsilon_m^2}{\lambda_{m0k}} 
\frac{J_0(\alpha_{0k}r/R) J_1(\alpha_{0k}\rho/R)}{\alpha_{0k} J_1^2(\alpha_{0k})} \cos(\pi m z/L) .
\end{eqnarray*}
\end{widetext}
Using the identity for $0 \leq x \leq 1$,
\begin{equation}
\sum\limits_{m=1}^\infty \frac{\cos(\pi m x)}{(\pi m)^2 + z^2} = \frac{\cosh z(1-x)}{2z\sinh z} - \frac{1}{2z^2} \,,
\end{equation}
one can evaluate the sum over $m$ that yields
\begin{eqnarray}  \nonumber
\hat{u}(r,z) &=& 1 - \frac{2\hat{j} \rho}{D} \sum\limits_{k} \frac{J_0(\alpha_{0k}r/R)J_1(\alpha_{0k}\rho/R)}{\alpha_{0k}^2 J_1^2(\alpha_{0k})} \\
\label{eq:u_SCA}
&& \times \frac{\cosh \alpha_{0k} (L-z)/R}{\sinh \alpha_{0k} L/R} .
\end{eqnarray}

The yet unknown flux density $\hat{j}$ is determined by imposing that
the solution $\hat{u}$ satisfies the Dirichlet boundary condition {\it
on average}, i.e.,
\begin{equation}  \label{eq:Dir_average}
0 = \int\limits_\Gamma d\x \, \hat{u}_{|z=0} \,,
\end{equation}
from which
\begin{equation}
\frac{\hat{j}}{D} = \frac{1}{4R} \biggl(\sum\limits_k \frac{J_1^2(\alpha_{0k}\rho/R)}{\alpha_{0k}^3 J_1^2(\alpha_{0k})} \, \ctanh(\alpha_{0k}L/R)\biggr)^{-1} .
\end{equation}
Multiplying this flux density by the area of the reactive region, one
determines the diffusive flux of particles in the modified problem:
\begin{equation}
\hat{J} = C_0 \pi \rho^2 \hat{j} .
\end{equation}
From this flux, one gets Eq. (\ref{eq:Psi_SCA}) for the normalized
flux $\Psi$.

In the case of imperfect reactions, one uses the same inhomogeneous
Neumann condition (\ref{eq:inhom}) to substitute the mixed
Robin-Neumann boundary conditions (\ref{eq:eq4}, \ref{eq:Robin}).  As
a consequence, the solution (\ref{eq:u_SCA}) of the modified problem
remains the same, whereas the average Dirichlet condition
(\ref{eq:Dir_average}) determining the effective flux density
$\hat{j}$ is replaced by
\begin{equation} 
0 = \int\limits_\Gamma d\x \, \biggl(\hat{u} - \frac{D}{\kappa} \partial_z \hat{u}\biggr)_{|z=0} ,
\end{equation}
from which
\begin{equation}
\frac{\hat{j}}{D} = \frac{1}{4R} \biggl(\frac{D}{4\kappa R} + 
\sum\limits_k \frac{J_1^2(\alpha_{0k}\rho/R)}{\alpha_{0k}^3 J_1^2(\alpha_{0k})} \, \ctanh(\alpha_{0k}L/R)\biggr)^{-1} 
\end{equation}
and thus 
\begin{equation}  \label{eq:Psi_sca_kappa}
\Psi_{\rm sca} = \frac12 \biggl(\frac{2D}{\kappa \rho} + \frac{8R}{\rho}
\sum\limits_k \frac{J_1^2(\alpha_{0k}\rho/R)}{\alpha_{0k}^3 J_1^2(\alpha_{0k})} \, \ctanh(\alpha_{0k}L/R)\biggr)^{-1} .
\end{equation}

\end{document}